\documentclass[preprint,eqsecnum,preprintnumbers,nofootinbib,byrevtex,prd,aps,showpacs,showkeys,groupedaddress,floatfix]{revtex4}
\usepackage{bm}
\usepackage{graphics}
\usepackage{graphicx}
\usepackage{epsfig}
\usepackage{amssymb}
\usepackage{amsmath}
\begin{document}

\title{Analytical solutions of the  Schr\"{o}dinger
equation   with the Woods-Saxon potential for arbitrary $l$ state}
\author{V.~H.~Badalov$^{1}$} \email{E-mail: badalovvatan@yahoo.com}
\author{H.~I.~Ahmadov$^{2}$}\email{E-mail: hikmatahmadov@yahoo.com}
\author{A.~I.~Ahmadov$^{1}$}\email{E-mail: ahmadovazar@yahoo.com}
\affiliation{$^{1}$Institute for Physical Problems,\\ Baku State
University, Z. Khalilov st. 23, AZ-1148, Baku,Azerbaijan\\
$^{2}$Department of Mathematical Physics,Faculty of Applied
Mathematics and Cybernetics\\
Baku State University, Z. Khalilov st. 23, AZ-1148, Baku,
Azerbaijan}\


\begin{abstract}
{In this work, the analytical solution of the radial Schr\"{o}dinger
equation for the Woods-Saxon potential is presented. In our
calculations, we have applied the Nikiforov-Uvarov method by using
the Pekeris approximation to the centrifugal potential for arbitrary
$l$ states. The bound state energy eigenvalues and corresponding
eigenfunctions are obtained for various values of $n$ and $l$
quantum numbers.}
\end{abstract}


\maketitle

\section{\bf Introduction}

An analytical solution of the radial Schr\"{o}dinger equation is of
high importance in nonrelativistic quantum mechanics, because the
wave function contains all necessary information for full
description of a quantum system. There are only few potentials for
which the radial Schr\"{o}dinger equation  can be solved explicitly
for all $n$ and $l$. So far, many methods were developed, such as
supersymmetry (SUSY) [1,2] and the Pekeris approximation [3-6], to
solve the radial Schr\"{o}dinger equation exactly or quasi-exactly
for $l\neq0$ within these potentials. Levai and Williams suggested a
simple method for constructing potentials for which the
Schr\"{o}dinger equation can be solved exactly in terms of special
functions [7] and showed relationship between the introduced
formalism  and supersymmetric quantum mechanics [1]. Amore \emph{et
al.} presented a new method [8] for the solution of the
Schr\"{o}dinger equation applicable to problems of nonperturbative
nature. In addition, they applied the method to the quantum
anharmonic oscillator and found energy eigenvalues and wave
functions, even for strong couplings.

The radial Schr\"{o}dinger equation for the Woods-Saxon potential
[9] can not be solved exactly for $l\neq0$. However, Fl\"{u}gge gave
an exact expression for the wave function, but a graphical method
was suggested for the energy eigenvalues at $l=0$ [4]. It is well
known that the Woods-Saxon potential is one of the important
short-range potentials in physics. Furthermore, this potential were
applied to numerous problems, in nuclear and particle physics,
atomic physics, condensed matter, and chemical physics.

Recently, an alternative method known as the Nikiforov-Uvarov (NU)
method [10] was proposed for solving the Schr\"{o}dinger equation.
Therefore, it would be interesting and important to solve the
nonrelativistic radial Schr\"{o}dinger equation for Woods-Saxon
potential for $l\neq0$, since it has been extensively used to
describe the bound and continuum states of the interacting systems.
Thus, one needs to obtain the energy eigenvalues and corresponding
eigenfunctions of the one particle problem within this potential.
The NU method was used by C. Berkdemir et al. [11] to solve the
radial Schr\"{o}dinger equation for the generalized Woods-Saxon
potential for $l=0$. However, it this work, the authors made errors
in application of the NU method, which led to no-correct results
[12].

In this work, we solve  the radial Schr\"{o}dinger equation for the
standard Woods-Saxon potential using NU method [10], and obtain the
energy eigenvalues and corresponding eigenfunctions for arbitrary
$l$ states.

\section{\bf Nikiforov-Uvarov method}

The Nikiforov-Uvarov (NU) method is based on the solutions of
general second-order linear equations with special orthogonal
functions. It has been extensively used to solve the nonrelativistic
Schr\"{o}dinger equation and other Schr\"{o}dinger-like equations.
The one-dimensional Schr\"{o}dinger equation or similar second-order
differential equations can be written using NU method in the
following form:
\begin{equation}
\psi''(z)+\frac{\widetilde{\tau}(z)}{\sigma(z)}{\psi}'(z)+
\frac{\widetilde{\sigma}(z)}{\sigma^{2}(z)}\psi(z)=0,
\end{equation}
where $\sigma(z)$ and $\widetilde{\sigma}(z)$ are polynomials, at
most second-degree, and $\widetilde{\tau}(z)$ is a first-degree
polynomial.

Using Eq.(2.1) the transformation
\begin{equation}
\psi(z)=\Phi(z)\\{y}(z)
\end{equation}
one reduces it to the hypergeometric-type equation
\begin{equation}
\sigma(z){y}''+\tau(z){y}'+\lambda{y}=0.
\end{equation}
The function $\Phi(z)$ is defined as the logarithmic derivative [10]
\begin{equation}
\frac{\Phi'(z)}{\Phi(z)}=\frac{\pi(z)}{\sigma(z)},
\end{equation}
where $\pi(z)$ is at most the first-degree polynomial.

The another part of $\psi(z)$, namely ${y}(z)$, is the
hypergeometric-type function, that for fixed $n$ is given by the
Rodriguez relation:
\begin{equation}
{y_{n}}(z)=\frac{{B_{n}}}{\rho(z)}\frac{{d^{n}}}{{dz^{n}}}[\sigma^{n}(z)\rho(z)],
\end{equation}
where ${B_{n}}$ is the normalization constant and the weight
function $\rho(z)$ must satisfy the condition [10]
\begin{equation}
\frac{d}{dz}\left(\sigma(z)\rho(z)\right)=\tau(z)\rho(z),
\end{equation}
with $\tau(z)=\widetilde{\tau}(z)+2\pi(z).$

For accomplishment of the conditions imposed on function $\rho(z)$,
the classical orthogonal polynomials, it is necessary, that
polynomial $\tau(z)$ becomes equal to zero in some point of an
interval $(a,b)$ and derivative of this polynomial for this interval
at $\sigma(z)>0$ will be negative, i.e. $\tau'(z)<0$.

The function $\pi(z)$ and the parameter $\lambda$ required for this
method are defined as follows:
\begin{equation}
\pi(z)=\frac{\sigma'-\widetilde{\tau}}{2}\pm\sqrt{\left(\frac{\sigma'-
\widetilde{\tau}}{2}\right)^{2}-\widetilde{\sigma}+{k}\sigma},
\end{equation}
\begin{equation}
\lambda=k+\pi'(z).
\end{equation}
On the other hand, in order to find the value of $k$, the expression
under the square root must be the square of a polynomial. This is
possible only if its discriminant is zero. Thus, the new eigenvalue
equation for the Schr\"{o}dinger equation becomes [10]
\begin{equation}
\lambda=\lambda_{n}=-n\tau'-\frac{n(n-1)}{2}\sigma'', (n=0,1,2,...).
\end{equation}
After the comparison of Eq.(2.8) with Eq.(2.9), we obtain the energy
eigenvalues.

\section{\bf Solutions of the Schr\"{o}dinger equation with the
Woods-Saxon potential}

The standard Woods-Saxon potential [9] is defined by
\begin{equation}
V(r)=-\frac{V_{0}}{1+\exp\left(\frac{r-R_{0}}{a}\right)}, a<<R_{0}.
\end{equation}
This potential was used for description of interaction of a neutron
with a heavy nucleus. The parameter $R_{0}$ is interpreted as radius
of a nucleus, the parameter $a$ characterizes thickness of the
superficial layer inside, which the potential falls from value $V=0$
outside of a nucleus up to value $V=-V_{0}$ inside a nucleus. At
$a=0$, one gets the simple potential well with jump of potential on
the surface of a nucleus.

The radial Schr\"{o}dinger equation [13, 14] with Woods-Saxon
potential is
\begin{equation}
\frac{d^2R}{dr^2}+\frac 2r\frac{dR}{dr}+\frac{2\mu}{\hbar^{2}}\left[
E+\frac{V_0 }{1+\exp \left( \frac{r-R_0}{a}\right)
}\right]R-\frac{l\left(l+1\right) }{r^2}R=0,  (0\leq r<\infty),
\end{equation}
where $l$ is the angular momentum quantum number and $\mu$ is the
reduced mass.

Introducing a new function
$$u(r)=rR(r),$$
Eq.(3.2) takes the form

\begin{equation}
\frac{d^2u(r)}{dr^2}+\frac{2\mu}{\hbar^2}\left[E+\frac{V_0}{1+\exp\left(\frac{r-R_0}{a}\right)}-
\frac{\hbar^{2}l(l+1)}{2\mu r^2}\right]u(r)=0.
\end{equation}
Equation (3.3) has the same form as the equation for a particle in
one dimension, except for two important differences. First, there is
a repulsive effective potential proportional to the eigenvalue of
$\hbar^{2}l(l+1)$. Second, the radial function must satisfy the
boundary conditions $u(0)=0$ and $u(\infty)=0.$

It is sometimes convenient to define in Eq.(3.3) the effective
potential in the form:
\begin{equation}
V_{eff}(r)=V(r)+\frac{\hbar^{2}l(l+1)}{2\mu r^{2}}.
\end{equation}

Then, the radial Schr\"{o}dinger equation  given by Eq.(3.3) takes
the form
\begin{equation}
\frac{d^2u}{dr^2}+\frac{2\mu }{\hbar^{2}} \left[
E-V_{eff}(r)\right]u=0.
\end{equation}

If in Eq.(3.1) introduce the notations
$$x=\frac{r-R_{0}}{R_{0}},\,\,\, \alpha=\frac{R_{o}}{a},$$
then the Woods-Saxon potential is given by the expression
$$V_{WS}=-\frac{V_{0}}{1+\exp(\alpha x)}.$$
The effective potential together with the WS potential for $l\neq0$
can be written as
\begin{equation}
V_{eff}(r)=V_{l}(r)+V_{WS}(r)=\frac{\hbar^{2}l(l+1)}{2\mu
r^{2}}-\frac{V_{0}}{1+\exp(\alpha x)}.
\end{equation}

It is known that the Schr\"{o}dinger equation cannot be solved
exactly for this potential at the value $l\neq0$ using the standard
methods as SUSY and NU. From Eq.(3.6) it is seen that the effective
potential is a combination of the exponential and inverse square
potentials, which cannot be solved analytically. Therefore, in order
to solve this problem we can take the most widely used and
convenient for our purposes Pekeris approximation. This
approximation is based on the expansion of the centrifugal barrier
in a series of exponentials depending on the internuclear distance,
taking into account terms up to second order, so that the effective
$l$-dependent potential preserves the original form. It should be
pointed out, however, that this approximation is valid only for low
vibrational energy states. By changing the coordinates
$x=\frac{r-R_{0}}{R_{0}}$ or $r=R_{0}(1+x)$, the centrifugal
potential is expanded in the Taylor series around the point $x=0$
($r=R_{0}$)
\begin{equation}
V_{l}(r)=\frac{\hbar^{2}l(l+1)}{2\mu r^{2}}=\frac{\hbar^{2}l(l+1)}
{2\mu
R_{0}^{2}}\frac{1}{(1+x)^{2}}=\delta\left(1-2x+3x^{2}-4x^{3}+\ldots\right),
\end{equation}
where $\delta=\frac{\hbar^{2}l(l+1)}{2\mu R_{0}^{2}}$.

According to the Pekeris approximation, we shall replace potential
$V_{l}(r)$ with expression

\begin{equation}
V^{*}_{l}(r)=\delta \left(C_{0}+\frac{C_{1}}{1+\exp\alpha
x}+\frac{C_{2}}{\left(1+\exp\alpha x\right)^{2}}\right).
\end{equation}

In order to define the constants $C_{0}$, $C_{1}$ and $C_{2}$, we
also expand this potential in the Taylor series around the point
$x=0$ ($r=R_{0}$):
\begin{equation}
V^{*}_{l}(x)=\delta\left[\left(C_{0}+\frac{C_{1}}{2}+\frac{C_{2}}{4}\right)-
\frac{\alpha}{4}\left(C_{1}+C_{2}\right)x+\frac{\alpha^{2}}{16}C_{2}x^{2}+\frac{\alpha^{3}
}{48}\left(C_{1}+C_{2}\right)x^{3}-\frac{\alpha^{4}}{96}C_{2}x^{4}+\cdots\right].
\end{equation}
Comparing equal powers of $x$ Eqs.(3.7) and (3.9), we obtain the
constants $C_{0}$, $C_{1}$ and $C_{2}$:
$$C_{0}=1-\frac{4}{\alpha}+\frac{12}{\alpha^{2}},\,\,
C_{1}=\frac{8}{\alpha}-\frac{48}{\alpha^{2}},\,\,
C_{2}=\frac{48}{\alpha^{2}}.$$

Now, the effective potential after Pekeris approximation becomes
equal to
\begin{equation}
V^{*}_{eff}(x)=V^{*}_{l}(x)+V_{WS}(x)=\delta
C_{0}-\frac{V_{0}-\delta C_{1}}{1+\exp(\alpha x)}+ \frac{\delta
C_{2}}{\left(1+\exp(\alpha x)\right)^{2}}.
\end{equation}

Instead of solving the radial Schr\"{o}dinger equation for the
effective Woods-Saxon potential $V_{eff}(r)$  given by Eq.(3.6), we
now solve the radial Schr\"{o}dinger equation for the new effective
potential $V^{*}_{eff}(r)$ given by Eq.(3.9) obtained using the
Pekeris approximation. Having inserted this new effective potential
into Eq.(3.5), we obtain
\begin{equation}
\frac{d^2u}{dr^2}+\frac{2\mu }{\hbar^{2}} \left[ E-\delta
C_{0}+\frac{V_{0}-\delta
C_{1}}{1+\exp\left(\frac{r-R_{0}}{a}\right)}-\frac{\delta
C_{2}}{\left(1+\exp\left(\frac{r-R_{0}}{a}\right)\right)^{2}}\right]u=0.
\end{equation}
We use the following dimensionless notations:
\begin{equation}
\epsilon^2=-\frac{2\mu \left(E-\delta C_{0}\right)a^2}{\hbar
^2};\,\,\beta^2=\frac{2\mu \left(V_{0}-\delta C_{1}\right)a^2}{\hbar
^2};\,\, \gamma^{2}=\frac{2\mu\delta C_{2}a^2}{\hbar^2},
\end{equation}
with  real $\epsilon>0$ for bound states; $\beta$ and $\gamma$ are
real and positive.

If we rewrite Eq.(3.11) by using a new variable of the form
$$z=\left(1+\exp\left(\frac{r-R_{0}}{a}\right)\right)^{-1},$$ we obtain
\begin{equation}
u^{\prime \prime }(z)+\frac{1-2z}{z(1-z)}u^{\prime }(z)+
\frac{-\epsilon^{2}+\beta^{2}z-\gamma^{2}z^{2}}{(z(1-z))^{2}}u(z)=0,
(0\leq z\leq 1),
\end{equation}
with $\widetilde{\tau}(z)=1-2z;  \sigma (z)=z(1-z);
\widetilde{\sigma}(z)=-\epsilon^2+\beta^2z-\gamma^{2}z^{2}$.

In the NU-method, the new function $\pi(z)$ is
\begin{equation}
\pi(z)=\pm\sqrt{\epsilon^2+\left(k-\beta^2\right)z-\left(k-\gamma^{2}\right)z^2}.
\end{equation}

The constant parameter $k$ can be found employing the condition that
the expression under the square root has a double zero, i.e., its
discriminant is equal to zero. Hence, there are two possible
functions for each $k$:
\begin{equation}
\label{2}\pi(z)=\pm\left\{
\begin{array}{c}
\left(\epsilon-\sqrt{\epsilon^{2}-\beta^{2}+\gamma^{2}}\right)z-\epsilon,\,\,\,
for \,\,\,
k=\beta^{2}-2\epsilon^{2}+2\epsilon\sqrt{\epsilon^{2}-\beta^{2}+\gamma^{2}},\\
\left(\epsilon+\sqrt{\epsilon^{2}-\beta^{2}+\gamma^{2}}\right)z-\epsilon,\,\,\,
for \,\,\,\,
k=\beta^{2}-2\epsilon^{2}-2\epsilon\sqrt{\epsilon^{2}-\beta^{2}+\gamma^{2}}.\,
\end{array}
\right.
\end{equation}
According to the NU-method, from the four possible forms of the
polynomial $\pi(z)$ we select the one for which the function
$\tau(z)$ has the negative derivative and root lies in the interval
(0,1). Therefore, the appropriate functions $\pi(z)$ and $\tau(z)$
have the following forms:
\begin{equation}
\pi(z)=\epsilon-\left(\epsilon+\sqrt{\epsilon^{2}-\beta^{2}+\gamma^{2}}\right)z,
\end{equation}

\begin{equation}
\tau(z)=1+2\epsilon-2\left(1+\epsilon+\sqrt{\epsilon^{2}-\beta^{2}+\gamma^{2}}\right)z,
\end{equation}
and
\begin{equation}
k=\beta^{2}-2\epsilon^{2}-2\epsilon\sqrt{\epsilon^{2}-\beta^{2}+\gamma^{2}}.
\end{equation}
Then, the constant $\lambda=k+\pi'(z)$ is written as
\begin{equation}
\lambda=\beta^{2}-2\epsilon^{2}-2\epsilon\sqrt{\epsilon^{2}-\beta^{2}+\gamma^{2}}-
\epsilon+\sqrt{\epsilon^{2}-\beta^{2}+\gamma^{2}}.
\end{equation}
An alternative definition of $\lambda_n$ (Eq.(2.9)) is
\begin{equation}
\lambda=\lambda_n=2\left(\epsilon+\sqrt{\epsilon^{2}-
\beta^{2}+\gamma^{2}}\right)n+n(n+1).
\end{equation}
Having compared Eqs.(3.19) and (3.20)
\begin{equation}
\beta^{2}-2\epsilon^{2}-2\epsilon\sqrt{\epsilon^{2}-\beta^{2}+\gamma^{2}}-
\epsilon-\sqrt{\epsilon^{2}-\beta^{2}+\gamma^{2}}=2\left(\epsilon+\sqrt{\epsilon^{2}-
\beta^{2}+\gamma^{2}}\right)n+n(n+1),
\end{equation}
we obtain
\begin{equation}
\epsilon+\sqrt{\epsilon^{2}-\beta^{2}+\gamma^{2}}+
n+\frac{1}{2}-\frac{1}{2}\sqrt{1+4\gamma^{2}}=0,
\end{equation}
or
\begin{equation}
\epsilon+\sqrt{\epsilon^{2}-\beta^{2}+\gamma^{2}}- n'=0.
\end{equation}
Here
\begin{equation}
n'=-n+\frac{\sqrt{1+4\gamma^{2}}-1}{2},
\end{equation}
$n$ being the radial quantum number $(n=0,1,2,\ldots)$. From
Eq.(3.23), we find
\begin{equation}
\epsilon=\frac{1}{2}\left(n'+\frac{\beta^{2}-\gamma^{2}}{n'}\right).
\end{equation}
Because for the bound states $\epsilon>0$, we get
\begin{equation}
n'>0
\end{equation}
and
\begin{equation}
-n'^{2}<\beta^{2}-\gamma^{2}<n'^{2}.
\end{equation}
If $n'>0$, there exist bound states, otherwise, there are no bound
states at all. By using Eq.(3.24) this relation can be recast into
the form
\begin{equation}
0\leq n<\frac{\sqrt{1+4\gamma^{2}}-1}{2},
\end{equation}
i.e. it gives the finite coupling value.

If $-n'^{2}<\beta^{2}-\gamma^{2}<n'^{2}$, there exists bound states;
otherwise there are no bound states. Inequality, which obtained
after substituting $\beta, \gamma,C_{0}, C_{1}, C_{2}$ into
Eq.(3.27), gives the definite coupling value for the potential depth
$V_{0}$:
\begin{equation}
\frac{8\delta}{\alpha}-\frac{\hbar^2}{2\mu a^2}n'^{2}<
V_{0}<\frac{8\delta}{\alpha}+\frac{\hbar^2}{2\mu a^2}n'^{2}
\end{equation}

After substituting $\alpha, \gamma, \delta, C_{2}$ and Eq.(3.24)
into Eqs.(3.28),(3.29), we find
\begin{equation}
0\leq n<\frac{\sqrt{1+\frac{192a^4l(l+1)}{R_{0}^4}}-1}{2}
\end{equation}
and
$$\frac{4\hbar^2al(l+1)}{\mu R_{0}^3}-\frac{\hbar^2}{8\mu
a^2}\left(\sqrt{1+\frac{192a^4l(l+1)}{R_{0}^4}}-2n-1\right)^{2}<$$
\begin{equation}
V_{0}<\frac{4\hbar^2 al(l+1)}{\mu R_{0}^3}+\frac{\hbar^2}{8\mu
a^2}\left(\sqrt{1+\frac{192a^4l(l+1)}{R_{0}^4}}-2n-1\right)^{2}
\end{equation}
or
$$\frac{4\hbar^2l(l+1)}{\mu }\frac{a}{R_{0}}-\frac{\hbar^2}{8\mu
}\frac{R_{0}^{2}}{a^{2}}\left(\sqrt{1+\frac{192a^4l(l+1)}{R_{0}^4}}-2n-1\right)^{2}<$$
\begin{equation}
V_{0}R_{0}^2<\frac{4\hbar^2l(l+1)}{\mu
}\frac{a}{R_{0}}+\frac{\hbar^2}{8\mu
}\frac{R_{0}^{2}}{a^{2}}\left(\sqrt{1+\frac{192a^4l(l+1)}{R_{0}^4}}-2n-1\right)^{2}.
\end{equation}
By defining
\begin{equation}
V_{0min}=\frac{4\hbar^2al(l+1)}{\mu R_{0}^3}-\frac{\hbar^2}{8\mu
a^2}\left(\sqrt{1+\frac{192a^4l(l+1)}{R_{0}^4}}-2n-1\right)^{2}
\end{equation}
and
\begin{equation}
V_{0max}=\frac{4\hbar^2 al(l+1)}{\mu R_{0}^3}+\frac{\hbar^2}{8\mu
a^2}\left(\sqrt{1+\frac{192a^4l(l+1)}{R_{0}^4}}-2n-1\right)^{2}
\end{equation}
Eq.(3.31) takes the following form:
\begin{equation}
V_{0min}< V_{0}<V_{0max}
\end{equation}
 The exact energy eigenvalues of the
Schr\"{o}dinger equation with the Woods-Saxon potential are derived
as
\begin{equation}
E_{nl}=\delta C_{0}-\left(V_{0}-\delta
C_{1}\right)\left(\frac{n'^2+\beta^2-\gamma^{2}}{2\beta
n'}\right)^{2}.
\end{equation}
Substituting the values of $\delta, C_{0}, C_{1}, C_{2}, n', \beta$
and $\gamma$ into (3.36), one can find $E_{nl}$
$$E_{nl}=\frac{\hbar^2 l(l+1)}{2\mu
R_{0}^2}\left(1+\frac{12a^{2}}{R_{0}^2}\right)-$$
\begin{equation}
\frac{\hbar^2}{2\mu
a^2}\left\{\frac{\left[\sqrt{1+\frac{192l(l+1)a^{4}}{R_{0}^4}}-2n-1\right]^2}{16}+
\frac{4\left[\frac{\mu a^2V_{0}}{\hbar^2}
-\frac{4l(l+1)a^{3}}{R_{0}^3}\right]^2}{\left[\sqrt{1+\frac{192l(l+1)a^{4}}{R_{0}^4}}-2n-1\right]^2}+\frac{\mu
V_{0}a^2}{\hbar ^2}\right\}
\end{equation}

If both conditions (3.28) and (3.29) ((3.30) and (3.31)) are
satisfied simultaneously, the bound states exist. Thus, the energy
spectrum equation (3.37) is limited, i.e. we have only the finite
number of energy eigenvalues.

For very large $V_{0}$  the $l$-dependent effective potential has
the same form as the potential with $l=0$. From Eqs.(3.30) and
(3.31), it is seen that if $l=0$, then one gets $n<0$. Hence, the
Schr\"{o}dinger equation for the standard Woods-Saxon potential with
zero angular momentum  has no bound state. For larger values of
$V_{0}$ ($V_{0}>V_{0max}$), the conditions (3.30) and (3.31) are not
satisfied. Therefore, no bound states exist for larger values of
$V_{0}$ ($V_{0}>V_{0max}$). Similarly, no bound states exist for
smaller values of $V_{0}$ ($0<V_{0}<V_{0min}$) too.

According to Eq.(3.37), the energy eigenvalues  depend on the depth
of the potential $V_{0}$, the width of the potential $R_{0}$, and
surface thickness $a$. Any energy eigenvalue must be less than
$V_{0}$. If constraints imposed on $n$ and $V_{0}$ satisfied, the
bound states appear. From Eq.(3.31), it is seen that the potential
depth increases when the parameter $R_{0}$ increases, but the
parameter $a$ is declined and vice versa. Therefore, one can say
that the bound states exist within this potential.

By using the empirical values $r_{0}=1.285fm$ and $a=0.65 fm$ taken
from Ref.15, the potential depth $V_{0}=(40.5+0.13A)MeV = 47.78 MeV$
and the radius of the nucleus $R_{0}=r_{0}A^{1/3}=4.9162 fm$ are
calculated for the atomic mass number of target nucleus $A = 56$ and
the reduced mass $\mu= 0.50433 u$. In Table 1, energies of the bound
states obtained numerically for the spherical standard Woods-Saxon
potential for some  values of $l$ and $n$ are given. It is clearly
seen from Table 1 that the $V_{0min}$ and $E_{nl}$ increase, but
$V_{0max}$ reduces when for the fixed values $l$ the parameter $n$
is increased. Therefore, $\triangle V=V_{0max}-V_{0min}$ reduces,
when $n$ increases.

In addition, we have seen that there are some restrictions on the
potential parameters for the bound state solutions within the
framework of quantum mechanics. That is, when the values of the
parameters $V_{0}$ and $n$ satisfy the conditions (3.30) and (3.31),
we obtain the bound states. We also point out that the exact results
obtained for the standard Woods-Saxon potential may have some
interesting applications for studying different quantum mechanical
and nuclear scattering problems. Consequently, the found wave
functions are physical ones.

Now, we are going to determine the radial eigenfunctions of this
potential. Having substituted $\pi(z)$ and $\sigma(z)$ into Eq.(2.4)
and then solving first-order differential equation, one can find the
finite function $\Phi(z)$ in the interval $[0,1]$

\begin{equation}
\Phi(z)=z^{\epsilon}\left(1-z\right)^{\sqrt{\epsilon^{2}-\beta^{2}+\gamma^{2}}}.
\end{equation}

It is easy to find the second part of the wave function from the
definition of weight function:
\begin{equation}
\rho(z)=z^{2\epsilon}\left(1-z\right)^{2\sqrt{\epsilon^{2}-\beta^{2}+\gamma^{2}}},
\end{equation}
and substituting into Rodrigues relation (2.4), we get
\begin{equation}
y_{n}(z)=B_{n}z^{-2\epsilon}\left(1-z\right)^{-2\sqrt{\epsilon^{2}-\beta^{2}+\gamma^{2}}}
\frac{d^{n}}{dz^{n}}\left[z^{n+2\epsilon}\left(1-z\right)^{n+2\sqrt{\epsilon^{2}-\beta^{2}+\gamma^{2}}}\right],
\end{equation}
where $B_{n}$ is the normalization constant and its value is
$\frac{1}{n!}$ [16]. Then, $y_{n}$ is given by the Jacobi
polynomials:
$$y_{n}(z)=P_{n}^{\left(2\epsilon,2\sqrt{\epsilon^{2}-\beta^{2}+\gamma^{2}}
\right)}(1-2z),$$
where $$P_{n}^{(\alpha,\beta)}(1-2z)=\frac{1}{n!}
z^{-\alpha}\left(1-z\right)^{-\beta}\frac{d^{n}}{dz^{n}}
\left[z^{n+\alpha}\left(1-z\right)^{n+\beta}\right].$$

The corresponding $u_{nl}(z)$ radial wave functions are found to be
\begin{equation}
u_{nl}(z)=C_{nl}z^{\epsilon}\left(1-z\right)^{\sqrt{\epsilon^{2}-\beta^{2}+\gamma^{2}}}P_{n}^{\left
(2\epsilon,2\sqrt{\epsilon^{2}-\beta^{2}+\gamma^{2}}\right)}(1-2z),
\end{equation}
where $C_{nl}$ is a new normalization constant determined using
$\int_o^\infty[u_{nl}(r)]^2dr=1$.

\section{\bf Conclusion}

In this paper, we have analytically calculated energy eigenvalues of
the bound states and corresponding eigenfunctions in the new exactly
solvable Woods-Saxon potential. The energy eigenvalue expression for
Woods-Saxon potentials is given by Eq.(3.37). As it should be
expected (see Eq.(3.37)), for any given set of parameters $V_{0},
R_{0}$ and $a$, the energy levels of standard Woods-Saxon potential
are positive. We can conclude that our results are interesting not
only for pure theoretical physicist but also for experimental
physicist, because the results are exact and more general.

\newpage
\begin{table}[h]
\begin{center}
\begin{tabular}{|c|c|c|c|c|c|}\hline
$l$ & $n$ & $V_{0min}, MeV$ & $V_{0max}, MeV$ & $V_{0},
MeV$&$E_{nl}, MeV$\\
\hline
    1 &  0  &    3.5590 &    3.7191 &    3.6   &    2.3374 \\ \hline
    2 &  0  &   10.2654 &   11.5690 &   10.5   &    7.0068 \\ \hline
    3 &  0  &   19.5398 &   24.1289 &   20     &   14.0327 \\ \hline
    4 &  0  &   30.8571 &   41.9241 &   35     &   22.6509 \\ \hline
    5 &  0  &   43.8248 &   65.3471 &   47.78  &   34.7761 \\ \hline
    6 &  0  &   58.1722 &   94.6684 &   90     &   35.3171 \\ \hline
    7 &  0  &   73.7179 &  130.0696 &  120     &   46.5875 \\ \hline
    8 &  0  &   90.3395 &  171.6730 &  160     &   54.5035 \\ \hline
    9 &  0  &  107.9536 &  219.5620 &  200     &   67.4620 \\ \hline
   10 &  0  &  126.5020 &  273.7949 &  240     &   85.1164 \\ \hline
   11 &  0  &  145.9433 &  334.4129 &  270     &  102.6090 \\ \hline
   12 &  0  &  166.2474 &  401.4463 &  284     &  153.2491 \\ \hline
   12 &  1  &  282.9924 &  284.7013 &  284     &  182.4285 \\ \hline
   13 &  0  &  187.3992 &  474.9172 &  330     &  177.8142 \\ \hline
   13 &  1  &  326.8691 &  335.4402 &  330     &  212.6080 \\ \hline
   14 &  0  &  209.3607 &  554.8423 &  390     &  198.7520 \\ \hline
   14 &  1  &  371.7128 &  392.4902 &  390     &  237.9290 \\ \hline
   15 &  0  &  232.1402 &  641.2347 &  450     &  223.1066 \\ \hline
   15 &     &  417.4850 &  455.8899 &  450     &  267.3521 \\ \hline
   20 &  0  &  357.9160 & 1170.4901 &  764.2   &  390.3832 \\ \hline
   20 &  1  &  659.4146 &  868.9915 &  764.2   &  465.7578 \\ \hline
   20 &  2  &  764.1028 &  764.3034 &  764.2   &  491.9299 \\ \hline
   30 &  0  &  667.3430 & 2716.9849 & 1690     &  834.2010 \\ \hline
   30 &  1  & 1204.0683 & 2180.2595 & 1690     &  968.3811 \\ \hline
   30 &  2  & 1543.9833 & 1840.3445 & 1690     & 1053.3543 \\ \hline
   30 &  3  & 1687.0879 & 1697.2400 & 1690     & 1088.9078 \\ \hline

\end{tabular}
\end{center}

\end{table}

\newpage
\begin{table}[h]
\begin{center}
\begin{tabular}{|c|c|c|c|c|c|}\hline
 $l$ & $n$ & $V_{0min}, MeV$ & $V_{0max}, MeV$ & $V_{0},
MeV $&$E_{nl}, MeV$\\
\hline
    40 &  0 & 1052.7566 & 4915.3065 & 3000 & 1430.1249 \\ \hline
    40 &  1 & 1826.2400 & 4144.8220 & 3000 & 1623.4737 \\ \hline
    40 &  2 & 2402.9130 & 3565.1490 & 3000 & 1767.5873 \\ \hline
    40 &  3 & 2782.7757 & 3185.2864 & 3000 & 1862.3459 \\ \hline
    40 &  4 & 2965.8279 & 3002.2441 & 3000 & 1904.9234 \\ \hline
    50 &  0 & 1513.7067 & 7765.9020 & 4600 & 2225.1111 \\ \hline
    50 &  1 & 2524.5792 & 6755.0294 & 4600 & 2477.7687 \\ \hline
    50 &  2 & 3338.6413 & 5940.9673 & 4600 & 2681.1670 \\ \hline
    50 &  3 & 3955.8930 & 5323.7156 & 4600 & 2835.2052 \\ \hline
    50 &  4 & 4376.3344 & 4903.2743 & 4600 & 2939.3913 \\ \hline
    50 &  5 & 4599.9653 & 4679.6434 & 4600 & 2986.8600 \\ \hline
   100 &  0 & 4948.2206 &31806.3078 &18400 & 8461.6670 \\ \hline
   100 &  1 & 7148.9350 &29605.5934 &18400 & 9011.8467 \\ \hline
   100 &  2 & 9152.8390 &27601.6894 &18400 & 9512.8202 \\ \hline
   100 &  3 &10959.9356 &25794.5958 &18400 & 9964.5902 \\ \hline
   100 &  4 &12570.2158 &24184.3126 &18400 &10367.1561 \\ \hline
   100 &  5 &13983.6886 &22770.8398 &18400 &10720.5172 \\ \hline
   100 &  6 &15200.3509 &21554.1774 &18400 &11024.6715 \\ \hline
   100 &  7 &16220.2029 &20534.3254 &18400 &11279.6153 \\ \hline
   100 &  8 &17043.2445 &19711.2838 &18400 &11485.3387 \\ \hline
   100 &  9 &17669.4757 &19085.0526 &18400 &11641.8108 \\ \hline
   100 & 10 &18098.8965 &18655.6319 &18400 &11748.8843 \\ \hline
   100 & 11 &18331.5069 &18423.0215 &18400 &11804.6769 \\ \hline

\end{tabular}
\end{center}
\caption{Energies of the bound states for the Woods-Saxon potentials
for different values of $n, l$ calculated using Eqs.(3.30), (3.33),
(3.34) and (3.37).} \label{table1}
\end{table}


\newpage

\begin{thebibliography}{99}
\bibitem{1}  F. Cooper, A. Khare and U. Sukhatme, Phys. Rep. {\bf 251} (1995) 267.
\bibitem{2}  D. A. Morales, Chem. Phys. Lett. {\bf 394} (2004) 68.
\bibitem{3}  C.L. Pekeris, Phys. Rev. {\bf 45} (1934) 98.
\bibitem{4}  S. Fl\"{u}gge, Practical Quantum Mechanics, Vol. 1 (Springer, Berlin, 1994)
\bibitem{5}  O. Bayrak and I. Boztosun, J. Phys. A {\bf 39} (2006) 6955 (arxiv: nucl-th / 0604042 V1).
\bibitem{6}  O. Bayrak, G. Kocak and I. Boztosun, J. Phys. A: Math. Gen. {\bf 39} (2006) 11521 (arxiv: math-ph / 0609010 V1).
\bibitem{7}  G. Levai and B. W. Williams, J. Phys. A: Math. Gen. {\bf 26} (1993) 3301.
\bibitem{8}  P. Amore, A. Aranda and A. De Pace, J. Phys. A: Math. Gen. {\bf 37} (2004) 3515.
\bibitem{9}  R. D. Woods and D. S. Saxon, Phys. Rev.  {\bf 95} (1954) 577.
\bibitem{10} A. F. Nikiforov and V. B. Uvarov, Special Functions of Mathematical Physics (Basel, Birkh\"{a}user, 1988).
\bibitem{11} C. Berkdemir, A. Berkdemir and R. Sever, Phys. Rev. C {\bf 72} (2005) 027001.
\bibitem{12} Editorial Note to Phys. Rev. C {\bf 72}, (2005) 027001; Phys. Rev. C {\bf 74} ( 2006) 039902(E).
\bibitem{13} L. D. Landau and E. M. Lifshitz, Quantum Mechanics (London, Pergamon Press, 1958).
\bibitem{14} W. Greiner, Quantum Mechanics (Springer, Berlin, 2001).
\bibitem{15} C. M. Perey, F. G. Perey, J. K. Dickens and R. J. Silva, Phys.Rev. {\bf 175} (1968) 1460.
\bibitem{16} H. Bateman and A. Erdelyi, Higher Transcendental functions, Vol. 2 (McGraw-Hill, New York, 1953)


\end{thebibliography}
\end{document}